\documentclass[aps,prl,twocolumn,superscriptaddress,showpacs]{revtex4}

\usepackage{graphicx}% Include figure files
\usepackage{dcolumn}% Align table columns on decimal point
\usepackage{bm}% bold math
\begin{document}
\def \lafepo{LaFePO}
\def \Ba122{BaFe$_2$As$_2$}
\def \CoBa122{Ba(Fe$_{1-x}$Co$_x$)$_2$As$_2$}
\def \KBa122{Ba$_{1-x}$K$_x$Fe$_2$As$_2$}
\def \LSCO{La$_{2-x}$Sr$_x$CuO$_4$}
\def \Tc{$T_c$}
\def \Torth{$T_{ortho}$}
\def \TN{$T_N$}

\title{In-plane electronic anisotropy in underdoped Ba(Fe$_{1-x}$Co$_x$)$_2$As$_2$ revealed by detwinning in a magnetic field}



%\thanks{}
\author{Jiun-Haw Chu}
\affiliation{Department of Applied Physics and Geballe Laboratory for Advanced Materials, Stanford University, Stanford, California 94305, USA}
\affiliation{Stanford Institute of Energy and Materials Science, SLAC National Accelerator Laboratory, 2575 Sand Hill Road, Menlo Park 94025,California 94305, USA}
\author{James G. Analytis}
\affiliation{Department of Applied Physics and Geballe Laboratory for Advanced Materials, Stanford University, Stanford, California 94305, USA}
\affiliation{Stanford Institute of Energy and Materials Science, SLAC National Accelerator Laboratory, 2575 Sand Hill Road, Menlo Park 94025,California 94305, USA}
\author{David Press}
\affiliation{E. L. Ginzton Laboratory, Stanford University, Stanford, California 94305, USA}
\author{Kristiaan De Greve}
\affiliation{E. L. Ginzton Laboratory, Stanford University, Stanford, California 94305, USA}
\author{Thaddeus D. Ladd}
\affiliation{E. L. Ginzton Laboratory, Stanford University, Stanford, California 94305, USA}
\affiliation{National Institute of Informatics, Hitotsubashi 2-1-2, Chiyoda-ku, Tokyo 101-8403, Japan}
\author{Yoshihisa Yamamoto}
\affiliation{E. L. Ginzton Laboratory, Stanford University, Stanford, California 94305, USA}
\affiliation{National Institute of Informatics, Hitotsubashi 2-1-2, Chiyoda-ku, Tokyo 101-8403, Japan}
\author{Ian R. Fisher}
\affiliation{Department of Applied Physics and Geballe Laboratory for Advanced Materials, Stanford University, Stanford, California 94305, USA}
\affiliation{Stanford Institute of Energy and Materials Science, SLAC National Accelerator Laboratory, 2575 Sand Hill Road, Menlo Park 94025,California 94305, USA}


\date{\today}

\begin{abstract}
We present results of angle-dependent magnetoresistance measurements and direct optical images of underdoped Ba(Fe$_{1-x}$Co$_x$)$_2$As$_2$ which reveal partial detwinning by action of a 14T magnetic field. Driven by a substantial magneto-elastic coupling, this result provides evidence for an electronic origin of the lattice distortion in underdoped iron pnictides. The observed anisotropy in these partially detwinned samples implies a substantial in-plane electronic anisotropy in the broken symmetry state, with a smaller resistivity along the antiferromagnetic ordering direction.
\end{abstract}

%\keywords{}
\maketitle

High temperature superconductivity emerges in the proximity of an antiferromagnetic(AF) ground state in several closely related families of iron-pnictides\cite{Hosono}. Interestingly, the AF transition in most of these materials is either preceded by or coincidental with a structural transition, which lowers the system's symmetry from tetragonal to orthorhombic\cite{Neutron}. It is generally believed that insights into the physics of this antiferromagnetism will lead to a better understanding of the origin of high temperature superconductivity in this family of compounds. To fully understand the magnetic ground state, it is crucial to measure the intrinsic in-plane anisotropy. However, the formation of structural twins in the orthorhombic crystals\cite{Ma}\cite{Tanatar}, makes such a measurement very challenging, and one would ideally like to find a simple method to detwin these materials and hence reveal the underlying in-plane anisotropy.

In 2002, Lavrov and coworkers performed a series of magnetotransport measurements accompanied by direct optical imaging on lightly doped \LSCO\ which has an orthorhombic transition around 450K, showing that it can be detwinned by a 14T magnetic field \cite{Ando}. Encouraged by this result, we have explored the possibility of affecting the structural/magnetic domains of iron pnictides in a similar manner, choosing Co-doped \Ba122 as a starting point. By doping with cobalt the single structural/magnetic transition of the undoped parent compound splits into two: the system first undergoes a tetragonal to orthorhombic structural transition(\Torth), then enters a collinear AF state at a lower temperature(\TN)\cite{CoBa122}\cite{ni}\cite{Lester}\cite{pratt}\cite{christianson}. Homogeneous doping and large single crystals, combined with the small temperature difference between \Torth\ and \TN , makes this an ideal material for such study.

In this Letter, we show via a combination of angle-dependent magnetoresistance(MR) measurements and direct optical images how magnetic fields can be used to detwin underdoped \CoBa122 . Our observations imply a surprisingly large in-plane electronic anisotropy below $T_N$ with a smaller resistivity along the AF ordering direction ($\rho_a < \rho_b$). This result has important consequences for our understanding of the reconstructed Fermi surfaces of underdoped iron pnictides, and for the driving force behind the structural transition.  

\begin{figure}
\includegraphics[width=8.5cm]{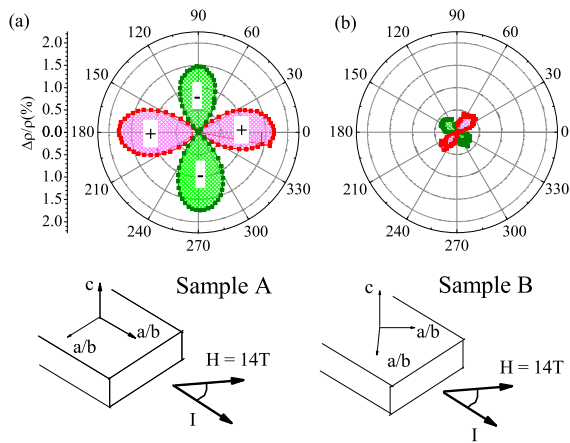}
\caption{\label{polar} (Color online) The in-plane magnetoresistance $\Delta\rho_{ab}/\rho_{ab}$ (\%) of \CoBa122\ for x = 2.5\% for two specific current orientations (sample A and B). The angle independent part of the magnetoresistance for sample B has been subtracted for clarity. The data were taken at 84K, below both \Torth\ and \TN\ for this cobalt concentration. The geometry of the measurement is indicated below each panel. Both current and field are in the ab-plane, and the angle of the magnetic field is measured relative to the current direction. For sample A current is applied parallel to the orthorhombic a/b axis, whereas for sample B current is applied at 45 degree to the orthorhombic a/b axis.}
\end{figure}

Single crystals of \CoBa122 were grown from a self flux, as described previously\cite{CoBa122}. Electrical contacts were made using sputtered gold pads in a standard four-point configuration, with typical contact resistance of 1-2 Ohms. Angle-dependent magnetotransport measurements were made in fields up to 14T. All the measurements were performed with both current and field parallel to the FeAs plane. Samples were cut into bar shapes of typical dimension 1mm$\times$0.2mm$\times$0.05mm and the crystal axes determined by x-ray diffraction. Except for the first set of experiments shown in Fig \ref{polar}, for which we deliberately varied the current orientation, all the measurements were taken with current along the orthorhombic a/b direction. Optical measurements were performed in a superconducting magnet cryostat, in which the field could be varied between 0 and 10T. The sample was illuminated with linearly polarized light, and viewed through an almost fully crossed polarizer in order to maximize the contrast of birefringence between neighboring domains. Crystals were mounted in an identical manner for both transport and optical measurements in such a way as to minimize external stress.

For the first set of experiments, samples of \CoBa122  were cut such that the direction of the current varied with respect to the crystal axes. Representative data are shown in fig. \ref{polar} for $x$=2.5\%  with current running along the orthorhombic [100] and [010] direction (sample A) and along the orthorhombic [110] direction (sample B). The configuration is illustrated schematically below each panel. At 84K, which is below both the structural and magnetic transitions for this cobalt concentration(\Torth = 99$\pm$0.5K and \TN = 93$\pm$0.5K respectively\cite{Lester}), we applied 14T and rotated the field within the FeAs plane to measure the resistivity as a function of angle. The angle-dependent MR of sample A is shown in the polar plot fig. \ref{polar}(a). It has a two-fold symmetry (as reported previously by Wang {\it et al.}\cite{XFWang}), with a positive MR for fields aligned parallel to the current, and negative when the field is perpendicular to the current. The angle dependent MR of sample B is shown in the polar plot fig. \ref{polar}(b). It also has a two-fold symmetry, but the magnitude is much smaller and the angle is shifted by 45 degrees. Clearly the two-fold MR is tied to the crystal axes and not to the current orientation.

\begin{figure}[h]
 \includegraphics[width=8.5cm]{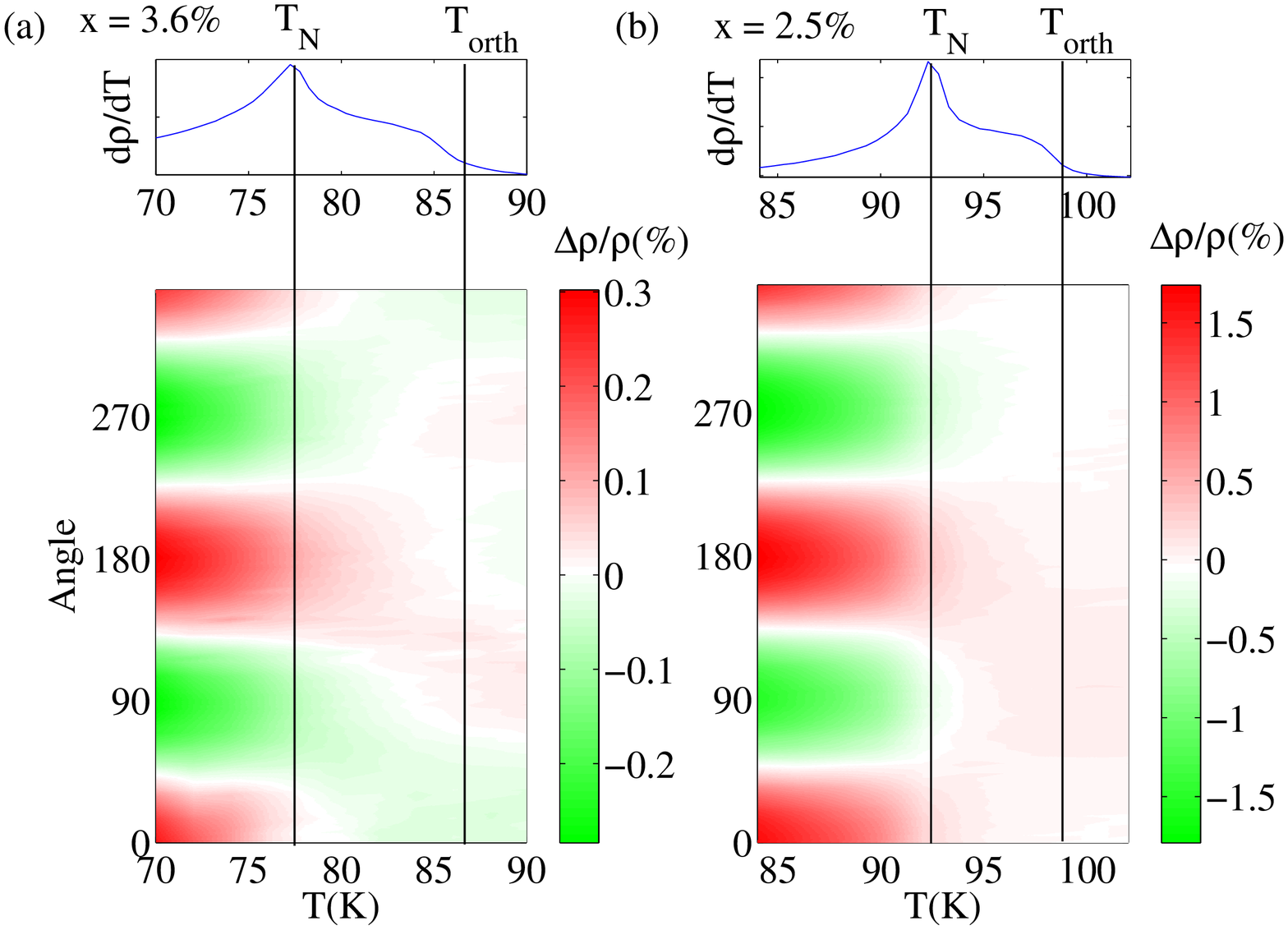}
\caption{\label{AMRO} (Color online) The angle dependent magnetoresistance $(\rho_{ab}(\theta,14T) - \overline{\rho_{ab}})/\overline{\rho_{ab}}$ (where $\overline{\rho_{ab}}$ is the average of $\rho_{ab}(\theta,14T)$ over $\theta$) as a function of angle and temperature of \CoBa122 for (a) $x$ =3.6\% and (b) $x$ = 2.5\%. The magnitude of MR is indicated by the color scale. Vertical lines indicate \Torth\ and \TN\, as determined for these samples by the derivative of the resistivity (shown above each panel).}
\end{figure}

To investigate the origin of this angle-dependent MR, we performed a detailed temperature dependent map close to the magnetic and structural transition temperatures. Representative data for $x$ = 2.5\% and 3.6\% , in a 14T magnetic field and using the ``Sample A" configuration ( i.e. current along orthorhombic a/b axis) are shown in Fig.\ref{AMRO}. These data are shown together with the temperature derivative of resistivity, which can be used to determine the two transition temperatures \cite{Lester}\cite{pratt}\cite{CoBa122}. The magnitude of the two-fold MR clearly rises sharply as the samples are cooled below \TN\ , although a very weak signal is visible up to almost \Torth\ for the higher cobalt concentrations.

\begin{figure}[tbh]
\includegraphics[width=8.5cm]{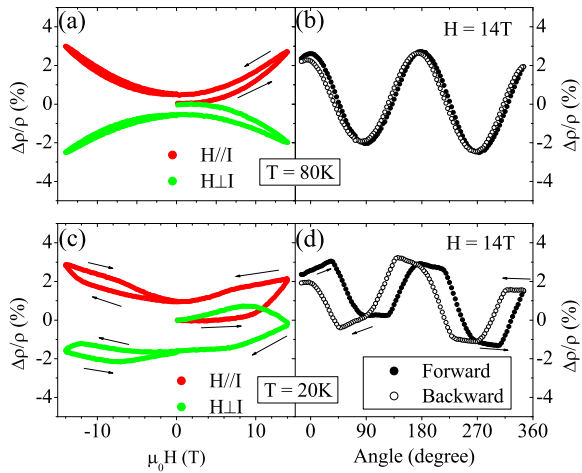}
\caption{\label{Field} (Color online) Representative magnetoresistance data for\CoBa122 as a function of field and angle for $x$ = 1.6\% at $T$ = 80K (panels (a) and (b)) and $T$ = 20K (panels (c) and (d)). For all cases the current is aligned parallel to orthorhombic $a$/$b$ axes and field is applied parallel to the $ab$ plane (i.e. sample A configuration). For field sweeps, the magnetic field was swept from 0 to 14T, then 14 to -14T, then back to 0T following an initial zero-field cool. Data were taken for fields aligned parallel (red) and perpendicular (green) to the current. Angle sweeps were performed in a field of 14T following an initial zero-field cool. Data were taken continuously as the angle was increased from 0 to 360 degrees and then back to zero.}
\end{figure}

To further investigate this effect, MR measurements were performed as a function of angle and field down to even lower temperatures. Representative data for $x$ = 1.6\% are shown in Fig. \ref{Field}. For temperatures close to \TN , the MR follows an almost perfect sinusoidal angle-dependence with minimal hysteresis as the angle between the field and the current is swept from 0 degrees to 360 degrees and back to 0 again (Fig. \ref{Field}(b)). The MR follows a $B^2$ field dependence for the entire field range, with a slight indication of some small hysteresis as the field is cycled back to zero (Fig. \ref{Field}(a)). Upon cooling to lower temperatures (for instance to 20K, as shown in panels (c) and (d) of Fig. \ref{Field}), the field dependence of the MR develops an apparent threshold behavior, with a threshold field close to 10T for the case shown in Fig. \ref{Field}(c). The MR also develops a substantial hysteresis. Both effects are also evident in the angle dependence of the MR (Fig. \ref{Field}(d)), which develops distinct shoulders, presumably related to the threshold behavior observed in the field dependence. The absolute value of the MR in 14T is comparable at the two temperatures shown. It is therefore unlikely that the predominant effect arises from orbital motion of the conduction electrons, which is ordinarily suppressed at higher temperatures according to Kohler’s rule \cite{Pippard}. Rather, the behavior shown in Fig. \ref{Field} is indicative of field-driven changes in the sample. These changes appear to be thermally assisted, such that the associated relaxation after the field is cycled to zero is much slower (essentially frozen on laboratory time scales) at lower temperatures, resulting in a larger hysteresis. The angle dependence implies that the projection of the applied field on to specific crystal orientations must exceed a specific threshold value to induce these changes at low temperatures. 

\begin{figure}[tbh]
\includegraphics[width=8.5cm]{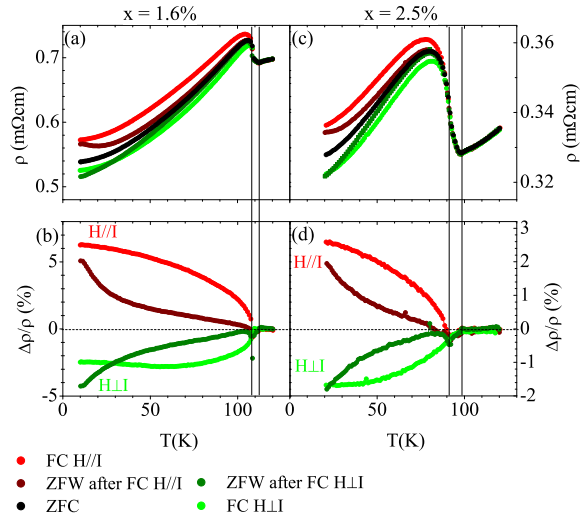}
\caption{\label{FCZFC} (Color online) Temperature dependence of the resistivity for representative samples with $x$ = 1.6\% (panels (a) and (b)) and $x$ = 2.5\% (panels (c) and (d)). Data were taken during an initial field cool (FC) in 14T, after which the field was cycled to zero and the resistivity measured in zero field while warming (ZFW). The current is aligned parallel to the a/b orthorhombic axes, and the field applied either parallel (red) or perpendicular (green) to the current. For comparison, data are also shown for the same samples while cooled in zero field (ZFC). The FC MR ($(\rho_{FC}-\rho_{ZFC})/\rho_{ZFC}$) and ZFW resistivity difference induced by FC ($(\rho_{ZFW}-\rho_{ZFC})/\rho_{ZFC}$) for the two field configurations are also plotted in (b) for x = 1.6\% and (d) for x = 2.5\% . }
\end{figure}

In order to maximize this effect, samples can be cooled through the structural/magnetic transitions in an applied magnetic field. Representative data are shown in Fig. \ref{FCZFC} for $x$ = 1.6\% and 2.5\% for fields applied both parallel and perpendicular to the current (still employing the “Sample A” configuration). Measurements were taken while cooling in an applied field of 14 T (field cool “FC”), and then while warming in zero field (zero field warm, “ZFW”) after cycling the field to zero at base temperature. The resistivity difference induced by field cooling along one orientation can be as large as 5\%\ at low temperature, even after the field is cycled to zero. The FC MR and ZFW resistivity difference for the two field configurations are also plotted in (b) and (d) for the two cobalt concentrations. The sign of MR for the two field orientations is opposite, and the absolute value appears to follow the magnetic order parameter, developing rapidly below \TN . A positive background in the FC MR data can be observed for both orientations, the magnitude of which increases as temperature is decreased, which is likely due to the ordinary metallic MR. In contrast, the resistivity difference of ZFW cycles appears to be rather symmetric. Its magnitude converges rapidly as the temperature is increased, consistent with thermally assisted relaxation, and with the hysteresis effect observed in field sweeps as a function of temperature.
																									
\begin{figure}[tbh]
\includegraphics[width=8.5cm]{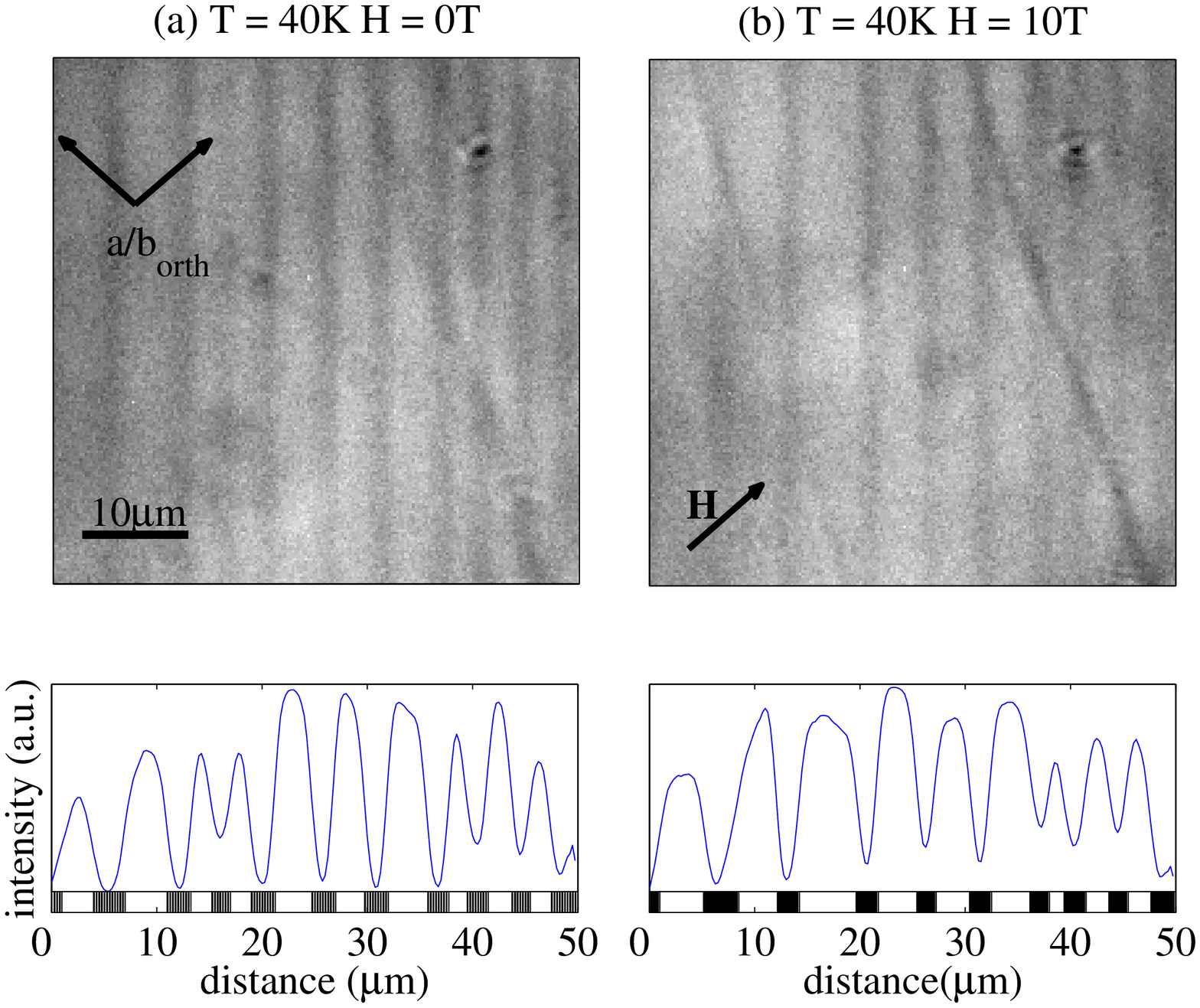}
\caption{\label{optical} (color online) Representative optical images of a \CoBa122 sample for x = 2.5\%. The images (obtained as described in the main text) were taken at $T$ = 40K below both \Torth\ and \TN . The initial image (a) was taken in zero field, following a zero field cool from above \TN . The field was then swept to 10T, at which field the second image (b) was taken. Horizontal intensity profiles, shown below each image, were calculated by integrating vertically over the image area after background subtraction and noise filtering. Boundaries between domains were estimated as described in the main text, resulting in an estimate of the relative fraction of the two domains indicated by black and white stripes below the figures. The field was oriented along the orthorhombic a/b axes.}
\end{figure}

We also performed direct optical measurements to observe the twinning domains on the surface of \CoBa122 . Representative images of a 2.5\% sample at $T$ = 40K are shown in Fig.\ref{optical}. Data were taken in zero field, and also after sweeping to 10 T. The presence of vertical stripes in both images indicates that the sample was twinned in zero and 10 T field. However, the relative density of the two domains is changed by application of the magnetic field. To track the difference of domain distributions, we plot the intensity profile across the boundaries, since the intensity depends on the crystal axes' orientation with respect to the light polarization. The twinning boundaries between two adjacent domains can then be determined from the maximum of the derivative of the intensity, from which we calculate the percentage of the volume of the domains of high intensity ($f = V_b/(V_a + V_b)$). For the two images shown in Fig.\ref{optical}, $f$ increases from 54 $\pm$ 1\%\ to 61 $\pm$ 1\%\, a difference $\Delta f$ = 7\% upon application of 10 T. Several regions in the optical images were analyzed, yielding percentage differences $\Delta f$ ranging from 5\% to 15\% . Despite the large range in $\Delta f$, which is due to the spatial variation of domain distribution, it is always positive, providing convincing evidence for partial field-induced detwinning.\footnote{An alternative explanation for the origin of the dark stripes in the images shown in Fig.\ref{optical} is that these regions consist of densely twinned domains. Such regions would be optically isotropic if the domain size is much smaller than the incident light's wavelength, giving rise to low intensity in the nearly crossed polarized configuration. Although we cannot rule out this possibility from the current measurements, nevertheless it is clear from the effect of magnetic field on the relative areas of light and dark regions in the optical images that the magnetic field detwins the sample.}

The origin of this detwinning effect is presumably related to the anisotropic in-plane susceptibility($\chi_a \neq \chi_b$) that must develop below \TN . Since the magnetic structure is collinear, with moments oriented along the long $a$-axis (referred to the orthorhombic unit cell)\cite{Neutron}, we can anticipate that $\chi_b > \chi_a$. In this case, fields oriented along the a/b axis of a twinned crystal will favor domains with the b-axis oriented along the field direction. Therefore for currents applied along the a/b axis and fields parallel to the current, the resistivity comprises a larger component of $\rho_b$ than $\rho_a$. For fields aligned perpendicular to the current the opposite is true. The effect clearly indicates the presence of a strong magneto-elastic coupling, suggestive of an electronic origin of the lattice distortion in these materials. Since the crystals are only partially detwinned, the observation of a positive MR for fields aligned parallel to the current implies a substantial anisotropy below \TN\ with $\rho_a < \rho_b$. \footnote{Partial detwinning reduces the twin boundary density, and therefore presumably decreases the contribution to the overall scattering rate from twin boundary scattering. However this effect cannot explain the permanent {\it increase} of resistivity for fields parallel to the current at low temperature. Moreover the twin boundaries extend along orthorhombic [110] direction\cite{Ma}\cite{Tanatar}, hence any scattering process would contribute equally to the resistivity along a and b-axis, inconsistent with the observed anisotropy}

This result has some important consequences. Intuitively one might expect to find that $\rho_a > \rho_b$, both because the a-axis lattice constant is larger than the b-axis lattice constant, and also because the spin-density wave(SDW) wave-vector is directed along the a-axis. Indeed, local-density approximation (LDA) calculations of the reconstructed Fermi surface for \Ba122 (incorporating a negative U to reduce the moment to the observed value)\cite{Analytis} indicate a larger plasma frequency for the b-axis relative to the a-axis\cite{Johannes}, consistent with this expectation in the limit of isotropic scattering. The observed anisotropy points towards either an unanticipated k-dependence of the scattering rate, or a stronger variation in the orbital character around the Fermi surface than is predicted by LDA.

In summary, we have shown how magnetic field can be used to detwin underdoped iron pnictides, opening a new avenue for research in to their anisotropic electronic properties. Our initial experiments have focused on the electron-doped system \CoBa122 , but given the generic collinear AF structure found in this family of compounds, it is likely that the effect is quite general, limited only by details of the twin-boundary pinning and the strength of field available. Our experiments reveal a surprisingly large in-plane anisotropy, with a smaller resistivity along the AF ordering direction. Understanding the source of this anisotropy will be a key step towards establishing the origin of superconductivity in these materials.

The authors thank C.-C. Chen, T.P. Devereaux and S.A. Kivelson for helpful discussions. We especially thank S.M. Hayden for initially suggesting that magnetic fields might be used to detwin iron pnictides. This work is supported by the DOE, Office of Basic Energy Sciences, under contract no. DE-AC02-76SF00515.

\end{document}